\documentclass[11pt,preprint]{aastex}

\def\kms  {km~s$^{-1}$}

\def\masy {mas~y$^{-1}$}

\def\uas  {\ifmmode {\mu{\rm as}}\else{$\mu$as}\fi}
\def\deg  {\ifmmode {^\circ}\else {$^\circ$}\fi}
\def\porm {\ifmmode {\pm}\else {$\pm$}\fi}
\def\chisqpdf {\ifmmode {\chi^2_{\rm pdf}}\else {$\chi^2_{\rm pdf}$}\fi}
\def\chisq    {\ifmmode {\chi^2}\else {$\chi^2$}\fi}

\def\etal {et al.~}
\def\eg   {e.g.,~}
\def\ie   {i.e.,~}

\def\d    {\ifmmode {{\rlap{.}}^\circ}\else {${\rlap{.}}^\circ$}\fi}
\def\s    {\ifmmode {{\rlap{.}}^s}\else {${\rlap{.}}^s$}\fi}
\def\as   {\ifmmode {{\rlap{.}}^{''}}\else {${\rlap{.}}^{''}$}\fi}

\newbox\grsign \setbox\grsign=\hbox{$>$} \newdimen\grdimen \grdimen=\ht\grsign
\newbox\laxbox \newbox\gaxbox
\setbox\gaxbox=\hbox{\raise.5ex\hbox{$>$}\llap
     {\lower.5ex\hbox{$\sim$}}}\ht1=\grdimen\dp1=0pt
\setbox\laxbox=\hbox{\raise.5ex\hbox{$<$}\llap
     {\lower.5ex\hbox{$\sim$}}}\ht2=\grdimen\dp2=0pt
\def\gax{\mathrel{\copy\gaxbox}}
\def\lax{\mathrel{\copy\laxbox}}

\def\pa    {\ifmmode {\psi} \else {$\psi$}\fi}
\def\rPpm  {\ifmmode {r_{\Ro,\To}} \else {$r_{Ro,\To}$}\fi}

\def\vlsr  {\ifmmode {V_{\rm LSR}}\else {$V_{\rm LSR}$}\fi}
\def\vlsrr {\ifmmode {v^r_{\rm LSR}}\else {$v^r_{\rm LSR}$}\fi}
\def\vhelio{\ifmmode {v_{Helio}}\else {$v_{Helio}$}\fi}

\def\ura   {\ifmmode {\mu_\alpha}\else {$\mu_\alpha$}\fi}
\def\udec  {\ifmmode {\mu_\delta}\else {$\mu_\delta$}\fi}

\def\ul    {\ifmmode {\mu_l}\else {$\mu_l$}\fi}
\def\ub    {\ifmmode {\mu_b}\else {$\mu_b$}\fi}

\def\uml   {\ifmmode {v_{gr}}\else {$v_{gr}$}\fi}
\def\umb   {\ifmmode {v_b}\else {$v_b$}\fi}
\def\vsrad {\ifmmode {v_{rad}}\else {$v_{rad}$}\fi}

\def\upl   {\ifmmode {v^p_{gr}}\else {$v^p_{gr}$}\fi}
\def\upb   {\ifmmode {v^p_b}\else {$v^p_b$}\fi}
\def\vprad {\ifmmode {v^p_{rad}}\else {$v^p_{rad}$}\fi}

\def\Vo    {\ifmmode {V^{Std}_\odot}\else {$V^{Std}_\odot$}\fi}
\def\Uo    {\ifmmode {U^{Std}_\odot}\else {$U^{Std}_\odot$}\fi}
\def\Wo    {\ifmmode {W^{Std}_\odot}\else {$W^{Std}_\odot$}\fi}
\def\VH    {\ifmmode {V^H_\odot}\else {$V^H_\odot$}\fi}
\def\UH    {\ifmmode {U^H_\odot}\else {$U^H_\odot$}\fi}
\def\WH    {\ifmmode {W^H_\odot}\else {$W^H_\odot$}\fi}
\def\V     {\ifmmode {V_\odot}\else {$V_\odot$}\fi}
\def\U     {\ifmmode {U_\odot}\else {$U_\odot$}\fi}
\def\W     {\ifmmode {W_\odot}\else {$W_\odot$}\fi}

\def\Vs    {\ifmmode {V_s}\else {$V_s$}\fi}
\def\Us    {\ifmmode {U_s}\else {$U_s$}\fi}
\def\Ws    {\ifmmode {W_s}\else {$W_s$}\fi}

\def\Vsbar {\ifmmode {\overline{V_s}}\else {$\overline{V_s}$}\fi}
\def\Usbar {\ifmmode {\overline{U_s}}\else {$\overline{U_s}$}\fi}
\def\Wsbar {\ifmmode {\overline{W_s}}\else {$\overline{W_s}$}\fi}

\def\aone  {\ifmmode {a_1}\else {$a_1$}\fi}
\def\atwo  {\ifmmode {a_2}\else {$a_2$}\fi}
\def\athr  {\ifmmode {a_3}\else {$a_3$}\fi}

\def\pars  {\ifmmode{\pi_s}\else{$\pi_s$}\fi}

\def\Ts    {\ifmmode{\Theta_s}\else{$\Theta_s$}\fi}
\def\Tdot  {\ifmmode{d\Theta\over dR}\else{$d\Theta\over dR$}\fi}

\def\Rp    {\ifmmode{R_p}\else{$R_p$}\fi}

\def\To    {\ifmmode{\Theta_0}\else{$\Theta_0$}\fi}
\def\Ro    {\ifmmode{R_0}\else{$R_0$}\fi}

\def\Dp    {\ifmmode{d_p}\else{$d_p$}\fi}
\def\Zsun  {\ifmmode {Z_\odot}\else {$Z_\odot$}\fi}
\def\Ytilt {\ifmmode{\psi_Y}\else{$\psi_Y$}\fi}
\def\Xtilt {\ifmmode{\psi_X}\else{$\psi_X$}\fi}

\def\ZIAU  {\ifmmode {Z_{IAU}}\else {$Z_{IAU}$}\fi}

\usepackage{listings}
\usepackage{appendix}

\usepackage{float}
\usepackage[caption = false]{subfig}
\usepackage{geometry}
\shorttitle{3D Kinematic Distances}
\shortauthors{Reid}

\begin{document}

\pagestyle{empty}

\title{\bf On the Accuracy of Three-dimensional Kinematic Distances}

\author{M. J. Reid\altaffilmark{1}}
\altaffiltext{1}{Center for Astrophysics~$\vert$~Harvard \& Smithsonian,
   60 Garden Street, Cambridge, MA 02138, USA}

\begin{abstract}
Over the past decade, the BeSSeL Survey and the VERA project have measured
trigonometric parallaxes to $\approx250$ massive, young stars using VLBI techniques.
These sources trace spiral arms over nearly half of the Milky Way.  What is now
needed are accurate distances to such stars which are well past the Galactic center.
Here we analyze the potential for addressing this need by combining line-of-sight
velocities and proper motions to yield 3D kinematic distance estimates.
For sources within about 10 kpc of the Sun, significant systematic uncertainties
can occur, and trigonometric parallaxes are generally superior.  However,
for sources well past the Galactic center, 3D kinematic distances are robust and more
accurate than can usually be achieved by trigonometic parallaxes. 
\end{abstract}

\section{Introduction}

Traditional kinematic distances for sources in the Galactic plane,
estimated by comparing line-of-sight (\vlsr) velocities
with those expected from a model of Galactic rotation, while extensively used for decades,
have significant limitations.  For sources within the ``Solar Circle'' (ie, Galactocentric
radii less than that of the Sun), a given \vlsr\ value occurs at two distances, and it
is often difficult to discriminate between them.  For Galactic longitudes
within about $\pm15\deg$ of the Galactic center and anticenter, $\vlsr$ values are near
zero for most distances and, thus, kinematic distances have very large uncertainties.
Also, near ``tangent points'' (where the
Sun--Source--Galactic Center angle is $90\deg$),
small uncertainties in \vlsr\ lead to large errors in distance,
since the gradient of \vlsr\ with distance vanishes.   Finally, with only one component
of a 3-dimenensional (3D) velocity vector, there is no consistency check on the
accuracy of the assumption of circular Galactic orbits.

\citet{Sofue:11} suggested using proper motions in addition to line-of-sight velocities
to constrain the Galaxy's rotation curve and, then, to estimate distances, emphasizing
the complementary information available in two components of motion.
Recently, \citet{Yamauchi:16} measured the proper motion in Galactic longitude
of the source G007.47+0.06 ($-5.03 \pm 0.07$ \masy).  They used this to obtain a
2-dimensional (2D) ``kinematic distance'' of $20\pm2$ kpc, corresponding to 12 kpc past the
Galactic center.   Shortly thereafter, \citet{Sanna:17} reported a trigonometric
parallax-distance of $20.4^{+2.8}_{-2.2}$ kpc for this source, supporting the
Yamauchi \etal\ estimate.

In this paper we expand on the 2D kinematic distance method by including proper motion
in Galactic latitude to obtain full 3D kinematic distances as described
in Section \ref{3D_method}.  We evaluate the accuracy of 3D kinematic distances
by comparing them to large numbers of parallax measurements in Section \ref{3d_vs_parallax}.
Next, we use simulations in Section \ref{simulations} to quantify expected distance
precision and accuracy in the presence of random non-circular motions.   Then, in
Section \ref{systematics}, we investigate sources of systematic error from inaccuracies
in the assumed Galactic rotation curve and for anomalous motions seen in sources near
the end of the Galactic bar.  Finally, in Section \ref{conclusions}, we summarize and
comment on potential applications of 3D kinematic distances.

\section{Estimating Distance with 3D Motions}\label{3D_method}

3D kinematic distance estimates come from combining probability density as a function of
distance for the line-of-sight velocity and the proper motion components in Galactic
longitude and latitude.   Assuming the measured values are Gaussianly distributed,
the probability density function (PDF) for the line-of-sight velocity component is
given by the following:
$${\rm Prob}_v(d|v,\sigma_v,RC) \propto {1\over\sigma_v} e^{-\Delta v^2/2\sigma_v^2}~~,\eqno(1)$$
where $\Delta v$ is the measured value of the line-of-sight velocity component minus
a value predicted by a rotation curve ($RC$), and $\sigma_v$ is the uncertainty in the measurement.
For one component of proper motion (toward Galactic longitude or latitude) the PDF is given by:
$${\rm Prob}_\mu(d|\mu,\sigma_\mu,RC) \propto {1\over\sigma_\mu}
                                  e^{-\Delta\mu^2/2\sigma_\mu^2}~~,\eqno(2)$$
where $\Delta\mu $ is the measured proper motion component minus
the velocity predicted from the rotation curve {\it divided by distance}.
For orbits within the Galactic plane, the expected Galactic latitude motion
should be small.  However, the PDF should be broader for nearby compared to distant sources,
thereby providing some distance information.

\begin{figure}[H]
\epsscale{0.77} 
\plotone{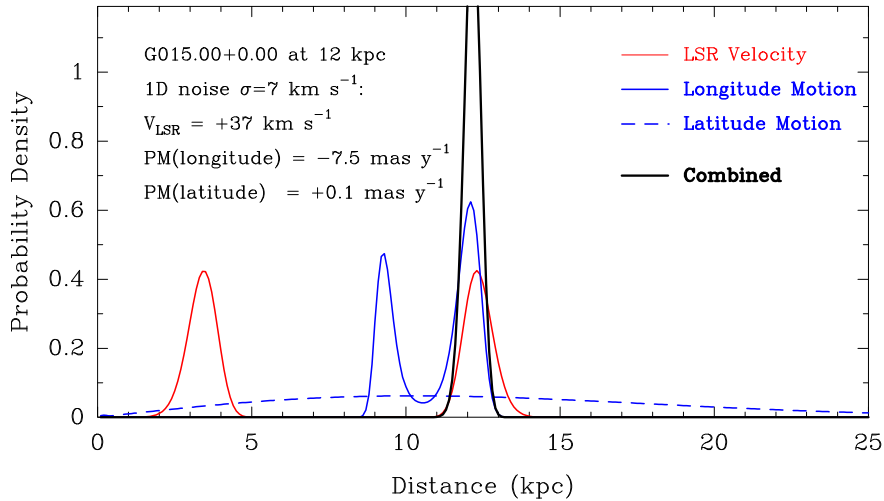}
\caption{\small Simulated probability density functions (PDFs) for 3 components of
  velocity for a source at longitude $15^\circ$ at a distance of 12 kpc and with 7 \kms\ noise
  added to each component (see listed \vlsr\ and proper motion (PM) values).
  The black likelihood function combines the 3 component PDFs (\ie\ 3D kinematic distance)
  and its peak and width provides an accurate and robust distance estimate and uncertainty.
       }
\label{fig:example}
\end{figure}

We multiply the three component PDFs to yield a combined 3D likelihood function, from which we
estimate a distance to the source from the location of peak of the function.  
Fig. \ref{fig:example} provides an example of the method for a simulated source
toward longitude $15\deg$ at 12 kpc distance and with 7 \kms\ random noise added
to each component of velocity.
Plotted are PDFs vs distance for each of the three components of
velocity, and the heavy black line is the combined 3D likelihood.
Note that the 3D likelihood resolves distance ambiguities apparent in both the
line-of-sight velocity and the longitude proper motion PDFs.  

We estimate distance from the peak of the combined 3D likelihood and assign a {\it formal}
uncertainty from the half-width in distance between points at $e^{-0.5}$ about the peak.
This formal uncertainty can underestimate a realistic error when the (unnormalized) 3D
likelihood peak is low.  This can happen, for example, when the line-of-sight velocity gives a
distance that is in significant tension with that from the Galactic longitude motion,
and the resulting 3D likelihood function can be too narrow to indicate uncertainty reliably.
In order to deal with such cases, we calculate an alternative uncertainty as half the
separation of the peaks of the line-of-sight velocity and Galactic longitude
motion PDFs.   We then report the maximum of this uncertainty and the formal
uncertainty to arrive at a more conservative distance uncertainty.

\section{Comparison of 3D Kinematic Distances with Parallaxes}\label{3d_vs_parallax}

\begin{figure}[H]
\epsscale{0.88} 
\plotone{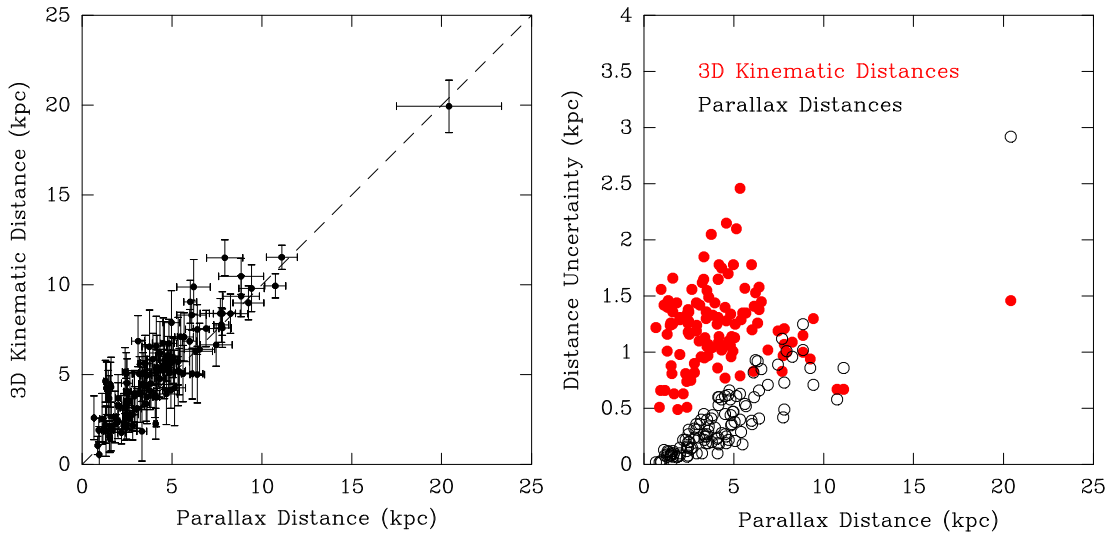}
\caption{\small VLBI parallax and proper motion data from \citet{Reid:19}
  selecting only sources with fractional parallax uncertainty $<15$\%.
  {\it Left panel:} Comparison of 3D kinematic vs parallax distances.
  {\it Right panel:} Uncertainties in the two methods vs distance.
  Note that 3D kinematic distance accuracy ({\it red dots)} matches and then
  surpases parallax accuracy ({\it black circles}) beyond about 8 kpc.
       }
\label{fig:3D}
\end{figure}

Using the parallaxes and proper motions compiled by \citet{Reid:19} from the Bar and
Spiral Structure Legacy (BeSSeL) Survey (bessel.vlbi-astrometry.org) and the
VLBI Exploration of Radio Astrometry (VERA) project (www.miz.nao.ac.jp),
the {\it left panel} of Fig. \ref{fig:3D} shows a good correspondence between the parallax
distances and the 3D kinematic distances.  
We note that 3D kinematic distances for sources nearby ($d\lax8$ kpc) are slightly biased
toward larger values compared to parallaxes.  This occurs because astrophysical
noise (eg, Virial motions of $\sim7$ \kms) in the longitude and latitude directions leads
to asymmetric PDFs when dividing by distance to convert to angular motion.
As distance approaches zero, this effect is magnified. Fortunately, this bias
decreases with increasing distance and does not significantly affect 3D kinematic
distances for sources past the Galactic center (see Section \ref{simulations} for
results of simulations).

\vskip 0.4truecm
The {\it right panel} of Fig. \ref{fig:3D}  shows measurement {\it uncertainty}
vs distance for the two methods.  For nearby sources, 3D kinematic distances are
considerably more uncertain than the parallax measurements.   However, a given
parallax uncertainty ($\sigma_\pi$) leads to increasing distance uncertainty with
distance (since $\sigma_d = d^2\sigma_\pi$), and beyond $\approx8$ kpc 3D kinematic
distances tend to be more precise.  The basic reason for this is that the longitude
proper motion can be very large.  For example, a source toward Galactic longitude 0.0 and
well past the Galactic center will have an apparent longitude speed of
$2\Theta_0 \approx 470$ km/s, since for circular orbits the Sun and the source are
moving in opposite directions with speeds of $\Theta_0$.  Even if one misestimates a
source's motion by 20 km/s (eg, by measuring only one water maser spot occurring in
an outflow of about that magnitude), that only corresponds to a 4\% error.
Also, errors of this magnitude in the assumed rotation curve will lead to similarly
small distance errors.  Except for sources within about 4 kpc of the Galactic center,
the rotation curve of \citet{Reid:19}, based on ``gold standard'' parallax
distances and measured 3D motions, should be accurate to about $\pm10$ \kms.  We conclude that
combining \vlsr\ and proper motion measurements can yield robust and precise
distances for sources well past the Galactic center.

\section{Effects of Random Non-Circular Motions}\label{simulations}

In this section, we investigate the accuracy of 3D kinematic distances by
simulating observations of sources whose motions differ from an assumed rotation curve
by the addition of random peculiar motions.   Specifically, for a given Galactic
longitude and distance from the Sun, we calculate the circular motion in the Galactic
plane from the ``universal rotation curve'' formulation of \citet{Persic:96}.  This
rotation curve is specified by only two parameters, and we adopt the $a2$ and $a3$ values
from \citet{Reid:19} (from model A5) obtained by fitting parallax and proper motion data
for 147 masers associated with very young ($<1$ My) and massive stars.

We add Gaussian random noise with $1\sigma$ dispersions, $\sigma_v$, of 7 or 20 \kms\
to each of
the three components of velocity.   A one-dimensional dispersion of 7 \kms\ corresponds
to a full magnitude dispersion of 12 \kms, and is a good approximation of random Virial
motions of young and massive stars within giant molecular clouds \citep{Reid:09}.  
A 20 \kms\ dispersion is representative of much older stars and is shown to indicate
the sensitivity of 3D kinematic distances to significantly larger dispersions.
For each simulated source, we calculate its probability density functions and estimate
distance as described in Section \ref{3D_method}.  

\begin{figure}[H]
\epsscale{0.77} 
\plotone{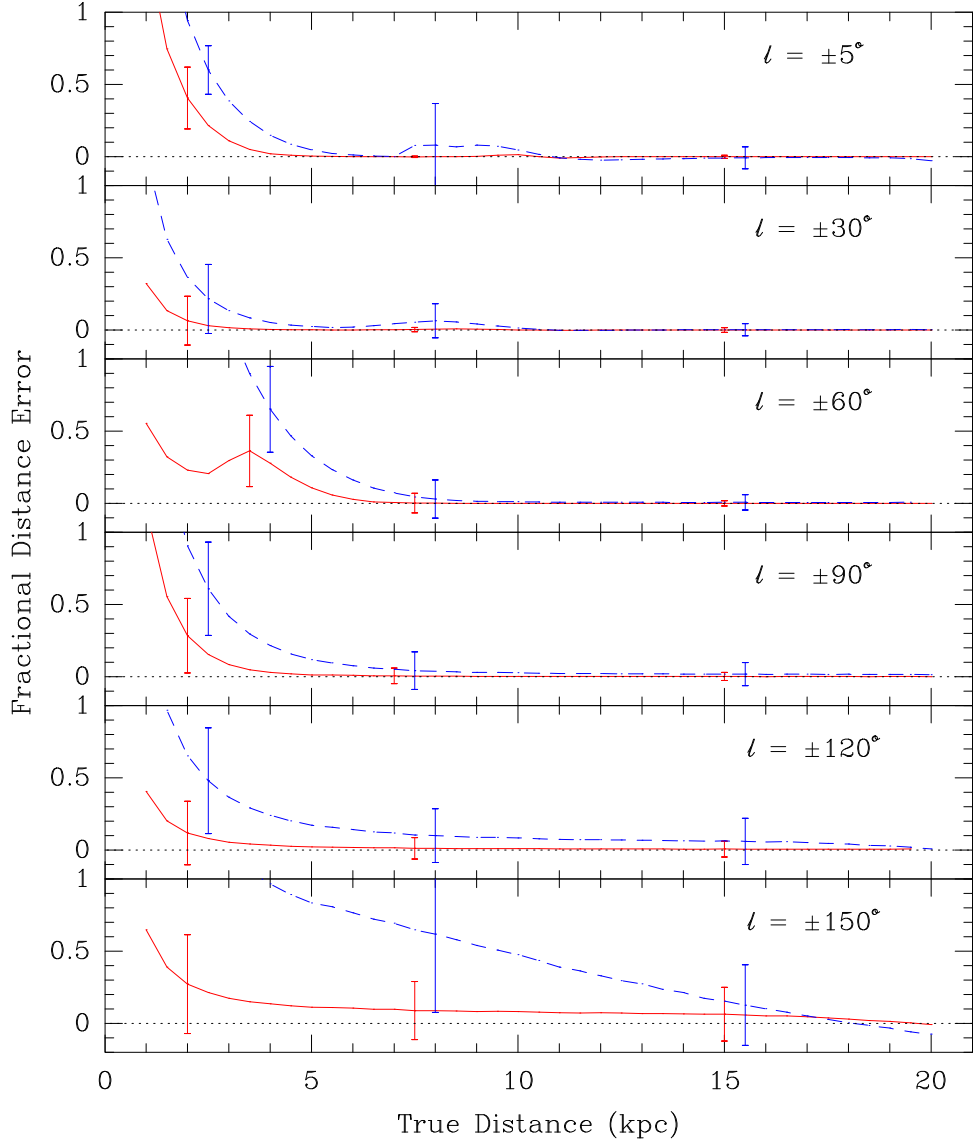}
\caption{\small Fractional distance error versus true distance for simulated data with
  7 \kms\ ({\it red lines}) and 20 \kms\ ({\it blue dashed lines}) random noise added.
  Lines are unweighted averages of 10,000 random trials sampled every 0.5 kpc;
  representative single-trial error bars indicate standard deviations about the mean.
  Galactic longitudes are indicated in the top right of each panel, with the $\pm$ symbol
  indicating that the same plot holds for both longitudes.
  Note the significant bias of 3D kinematic distance estimates, which grow toward the
  origin as the inverse of true distance.  See Section \ref{simulations} for more
  discussion.
       }
\label{fig:fractional_errors}
\end{figure}

Fig. \ref{fig:fractional_errors} displays the fractional distance error (ie,
the difference between simulated and true distances divided by the true distance)
as a function of the true distance for the two velocity dispersions.
We calculate 10,000 independent trials for each true distance sampled every 0.5 kpc
and connect the unweighted means of those trials.
The representative error bars in Fig. \ref{fig:fractional_errors} are standard
deviations about the mean value and indicate the expected statistical ($1\sigma$)
uncertainty in a single trial.  The simulations reveal both statistical uncertainties
and systematic biases in 3D kinematic distances.

Regarding statistical uncertainties, for magnitudes of Galactic longitude below
$\approx90\deg$ (ie, in Galactic quadrants I and IV) and true distances $\gax10$ kpc,
fractional distance uncertainties are $\lax10$\%, corresponding to $\lax\pm1$ kpc at 10 kpc.
For longitude magnitudes $>90\deg$ (ie, quadrants II and III), statistical
uncertainties start to grow, especially for the larger (20 \kms) random noise
used in the simulations.  This occurs because the sensitivity of both line-of-sight
and longitude motions, which are proportional gradients in circular motion with distance,
are low and random noise becomes relatively more important.

Systematic biases, owing to random peculiar motions, typically are less than several
percent of distance at most Galactic longitudes for true distances $\gax10$ kpc.
For example, at a longitude of $30\deg$ and a true distance of 12 kpc, random motions
of 20 \kms\ yield a bias in fractional distance of $<0.2$\%, and this bias only grows
to $1$\% at a longitude of $120\deg$.   However, a clear inverse-distance bias towards
larger estimated distances is seen for true distances $\lax5$ kpc.
This occurs because astrophysical noise (eg, Virial motions of $\sim7$ \kms)
in the longitude direction leads to an asymmetric PDF when dividing by distance
to convert to angular motion.  As distance approaches zero, this effect is magnifed.
Fortunately, this bias is predictable, provided the velocity noise is known, and,
in any event, it decreases with increasing distance and often becomes insignificant
for distances $\gax10$ kpc.   

A second systematic problem is evident, for example, in the $5\deg$ and $30\deg$
longitude plots in Fig. \ref{simulations} as a $\lax10$\% positive bias between true
distances of 6 to 10 kpc for 20 \kms\ added noise, and as a 30\% positive bias in
the $60\deg$ longitude plot near 3.5 kpc for 7 \kms\ added noise.  These cases can
also be identified by thier large standard deviations.  They are caused by a
combination of effects, including 1) near the tangent points (at 8.1, 7.1 and 4.1 kpc
for longitudes of $5\deg$, $30\deg$, and $60\deg$, respectively) the gradient
in the line-of-sight velocity vanishes, and 2) a small number of trials can have a
secondary probability peak at much larger distance which is slightly favored over the
``primary'' peak, owing to the addition of random motions.   These cases can be
recognized as problematic by examining the component PDFs in a manner similar to
Fig. \ref{fig:example}.  However, here we have not attempted to remove such
cases when presenting means and standard deviations from all trials.

\section{Effects of Systematic Errors}\label{systematics}

The 3D kinematic distance method requires accurate values for the Galaxy's fundamental
parameters: the distance to the Galactic center ($\Ro$), the circular rotation speed at
the Sun ($\To$), and its rotation curve.   Fortunately, the BeSSeL Survey and the VERA
project have provided hundreds of parallaxes and proper motions for massive and extremely
young ($<1$ My) stars, and these have been modeled to give accurate estimates of
$\Ro=8.15\pm0.15$ kpc and $\To=236\pm7$ \kms\ \citep{Reid:19}.
These values are independently confirmed by 1) infrared observations tracing stars orbiting
the supermassive black hole, Sgr A*, at the center of the Galaxy, giving
$\Ro=7.946\pm0.059$ kpc \citep{Do:19} and $\Ro=8.275\pm0.034$ kpc \citep{Gravity:21},
for which a variance-weighted average is $\Ro=8.19$ kpc, and 2) the apparent proper
motion in longitude of Sgr A* (giving the reflex of the Sun's orbital motion in the Galaxy)
of $6.411\pm0.008$ mas~y$^{-1}$ \citep{Reid:20}, which translates to
$\To=236\pm2$ \kms, after subtracting the Sun's peculiar motion of $12\pm2$ \kms\
\citep{Schoenrich:10} in the direction of Galactic rotation.

The BeSSeL Survey and VERA project measurements also, provide a ``gold standard'' rotation
curve, based on parallax distances and 3D velocities, between Galactocentric radii (R) of
4 to 15 kpc \citep{Reid:19}.  Thus, inaccuracies in the model of the Galaxy are not
likely to have a significant effect on 3D kinematic distances over this range of radii.
However, inside of 5 kpc the observational constraints for the adopted ``universal''
rotation curve model become weaker.  In order to assess the accuracies of 3D kinematic
distances in the central 5 kpc, we simulate uncertainty in the model rotation curve by adding
rms deviations which grow linearly from 0 \kms\ at $R=5$ kpc to 50 \kms\ at $R=0$ kpc.
When adding random speeds to the rotation curve values in simulated trials,
we do not allow rotation speed to drop below 50 \kms.

\begin{figure}[h]
\epsscale{0.77} 
\plotone{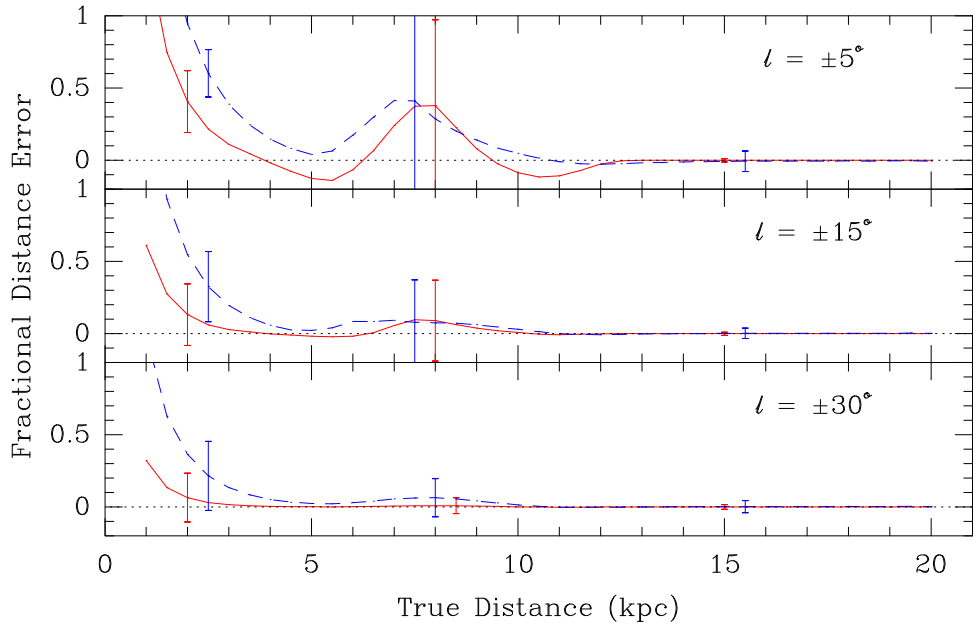}
\caption{\small Fractional distance error versus true distance for simulated data
  allowing for significant uncertainty in the assumed rotation curve speed within
  5 kpc of the Galactic center (at a distance of 8.15 kpc from the Sun) in addition to random motions.
  See Fig. \ref{fig:fractional_errors} caption for other details.
       }
\label{fig:galactic_center}
\end{figure}

Fig. \ref{fig:galactic_center} shows the effects of errors in the assumed
rotation curve for three Galactic longitudes for which the minimum Galactic
radii are less than 5 kpc.  At longitude $30\deg$ there is no noticeable
change in fractional distance uncertainty compared to that shown in Fig.
\ref{fig:fractional_errors}, since the minimum radius (at the tangent point
distance of 8.1 kpc) is 4.1 kpc, and so at no distances do we add more than
9 \kms\ noise to the trial rotation curves.  Noticeable effects are evident for
the longitude $15\deg$ plot, where the minimum Galactic radius is 2.1 kpc
and up to 29 \kms\ rms noise is added to the rotation curve.  However, at $5\deg$
longitude, where the minimum radius is only 0.7 kpc and up to 43 \kms\ noise
is added, there are significant fractional distance errors over a range of
distances from about 5 to 11 kpc.

\begin{figure}[h]
\epsscale{0.77} 
\plotone{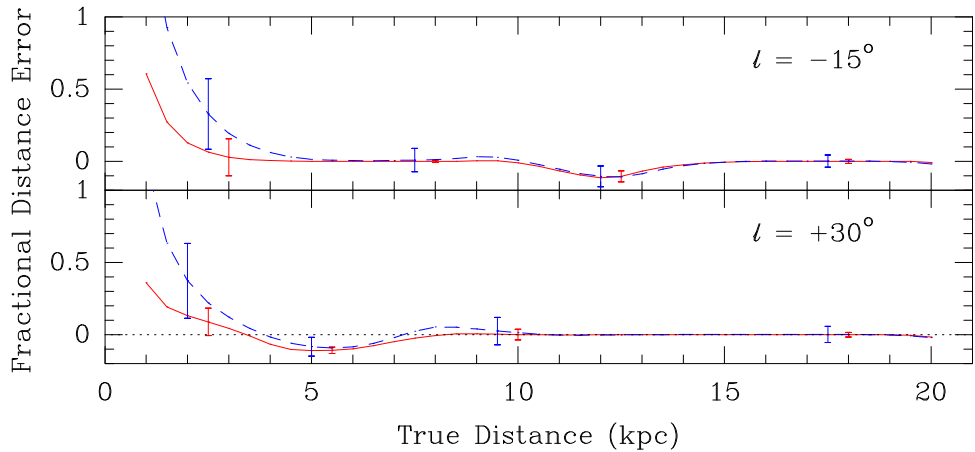}
\caption{\small Fractional distance error versus true distance for simulated data
  allowing for anomalous motions near the end of the Galactic bar toward the 
  Galactic center in addition to random motions.  The far end of the bar is near
  $-15\deg$ at 12 kpc ({\it top panel}).  The near end of the bar is near
  $+30\deg$ at 4 kpc ({\it bottom panel}).
  See Fig. \ref{fig:fractional_errors} caption for other details.
       }
\label{fig:galactic_bar}
\end{figure}

Fig. \ref{fig:galactic_bar} shows the effects of anomalous motions seen in
massive young stars near the end of the Galactic bar (Immer et al 2019).  These
anomalous motions peak at $\approx40$ \kms\ and are mostly directed toward the
Galactic center.  For the near end of the bar in Galactic quadrant 1, we
model these anomalies with a Gaussian distribution, centered at longitude
$27\deg$ and a Galactocentric radius of 4.5 kpc, and decreasing with offset
with a $1\sigma$ scale-length of 2 kpc.  The bottom panel of
Fig. \ref{fig:galactic_bar} indicates that for sources at Galactic longitude
$+30\deg$ and distances from the Sun of 4 to 7 kpc, one sees an $\approx10$\%
bias toward smaller values of 3D kinematic distances.
For the far end of the bar in Galactic quadrant 4, we also
model these anomalies with a Gaussian distribution, but centered at longitude
$-13\deg$ and a Galactocentric radius of 12 kpc.  The top panel of
Fig. \ref{fig:galactic_bar} indicates that for sources at Galactic longitude
$-15\deg$ and at distances from the Sun of 11 to 13 kpc, one again sees an
$\approx10$\% bias toward smaller values of 3D kinematic distances.
These bias are similar for the cases of 7 \kms\ ({\it solid red lines}) and
20 \kms\ ({\it dashed blue lines}) added noise, since the anomalous motions
are systematic and tend to dominate.  

\section{Conclusions and Outlook}\label{conclusions}

In this paper, we have examined the precision and accuracy of 3D kinematic distance
estimates in several ways.  First, we compared these estimates against trigonometric
parallax distances for large numbers of masers associated with massive young
stars.   These demonstrated a good corresponce between the two methods for
stars with distances $\gax8$ kpc, and indicated that 3D kinematic distances
can be more accurate than parallax measurments.  For more nearby stars, 3D
kinematic distances displayed a modest bias of $\approx1$ kpc toward larger distances.  

In order to better assess accuracies and biases for 3D kinematic distances,
we simulated large numbers of sources with (1D) random motions of 7 and 20 \kms\
and evaluated the statistical properties of such distances estimates.
Random motions of 7 \kms\ per velocity component (corresponding to 12 \kms\
for the full velocity magnitude) are representative of Viral motions of within
giant molecular clouds.  For these simulations, we find excellent performance
of 3D kinematic distances (ie, negligible bias and fractional distance uncertainty
less than 10\%) for true distances $\gax5$ kpc.  For smaller distances, a positive
bias of $\approx20$\% of the true distance is generally seen at 2 kpc true distance
and scales with the inverse of true distance.   Additional positive biases are seen
near tangent point distances of 4 to 8 kpc for Galactic longitudes of $60\deg$ and
$5\deg$, where line-of-sight velocity gradients are small and secondary peaks
in the distance PDFs can become favored.

We also used simulations to assess the effects of systematic errors in the
assumed rotation curve in the inner Galaxy, as well as for a region near the
end of the Galactic bar where large non-circular motions have been noted.
For distances near 8 kpc and Galactic longitudes $\lax15\deg$, uncertainty in
the rotation curve produces a noticeable increase in dispersion in distance
estimates, and at longitude $5\deg$ the dispersion becomes comparable to distance.  
Anamalous motions, as seen for massive young stars near the end of the Galactic bar
in Galactic quadrant 1, generally produce negative 3D kinematic distance biases of
$\lax10$\%.

Simulations assuming large random 1D velocities of 20 \kms\ (corresponding to
3D magnitudes of 35 \kms), as would be expected produce larger dispersions and
biases compared to the 7 \kms\ simulations.  However, for Galactic longitudes with
magnitudes $\lax120\deg$ and distances $\gax10$ kpc, 3D kinematic distances
still perform very well. 

3D kinematic distances have some obvious applications.  
Directly mapping the spiral structure of the Milky Way has 
proven to be a challenging enterprise, since distances are very large and 
dust extinction blocks most of the Galactic plane at optical wavelengths.  
Thus, {\it Gaia}, even with a parallax accuracy approaching $\pm0.01$ mas, cannot 
freely map the Galactic plane.  However, VLBI is unaffected by extinction
and can detect masers associated with young stars that best trace spiral
structure.  While a parallax with 12\% accuracy has been measured for a water
maser associated with a massive young star on the far side of the Milky
Way at 20 kpc distance, obtaining large numbers of such measurements would
require an enormous amount of observing time.  However, for such distant
massive, young stars (with random 1D motions near 7 \kms), 3D kinematic
distances have intrinsic accuracy better than can generally be achieved with
parallax measurements, and they can be obtained for large numbers of stars
with only modest amounts of observing time.  This will facilitate tracing 
spiral structure across the entire Milky Way.

Finally, we note that knowledge of distance to X-ray binaries is crucial
to understanding their nature.   For example, a trigonometric parallax-distance
measurement of 2.22 kpc for Cyg X-1 firmly established that this binary system contains
a black hole and a massive young star \citep{Miller-Jones:21}.  However,
accurate trigonometric parallax measurements for more distant binaries have proven
difficult to obtain, \eg\ GRS 1915 \citep{Reid:14}.  Since proper motions can be measured
far more easily than parallaxes, for distant binaries with modest
peculiar motions, 3D kinematic distances can provide reliable distances.

\vskip 0.2truecm

\end{document}